\documentclass[conference]{IEEEtran}
\IEEEoverridecommandlockouts
\usepackage{cite}
\usepackage{amsmath,amssymb,amsfonts}
\usepackage{algorithmic}
\usepackage{graphicx}
\usepackage{textcomp}
\usepackage{xcolor}
\def\BibTeX{{\rm B\kern-.05em{\sc i\kern-.025em b}\kern-.08em
    T\kern-.1667em\lower.7ex\hbox{E}\kern-.125emX}}
\begin{document}

\title{Improving Speech Inversion Through Self-Supervised Embeddings and Enhanced Tract Variables\\

\thanks{This work was supported by National Science Foundation grant 2141413.}
}

\author{\IEEEauthorblockN{1\textsuperscript{st} Ahmed Adel Attia}
\IEEEauthorblockA{\textit{Institute for Systems Research
} \\
\textit{University of Maryland, College Park}\\
Maryland, United States of America \\
aadel@umd.edu}
\and
\IEEEauthorblockN{1\textsuperscript{st}  Yashish M. Siriwardena}
\IEEEauthorblockA{\textit
{Institute for Systems Research
} \\
\textit{University of Maryland, College Park}\\
Maryland, United States of America \\
yashish@umd.edu}
\and
\IEEEauthorblockN{2\textsuperscript{nd} Carol Espy-Wilson}
\IEEEauthorblockA{\textit
{Institute for Systems Research
} \\
\textit{University of Maryland, College Park}\\
Maryland, United States of America \\
espy@umd.edu}}

\maketitle

\begin{abstract}
The performance of deep learning models depends significantly on their capacity to encode input features efficiently and decode them into meaningful outputs. Better input and output representation can potentially boost models’ performance and generalization. In the context of acoustic-to-articulatory speech inversion (SI) systems, we study the impact of utilizing speech representations acquired via self-supervised learning (SSL) models, such as HuBERT compared to conventional acoustic features. Additionally, we investigate the incorporation of novel tract variables (TVs) through an improved geometric transformation model. By combining these two approaches, we improve the Pearson Product Moment Correlation (PPMC) scores which evaluate the accuracy of TV estimation of the SI system from 0.7452 to 0.8141, a 6.9\% increase. Our findings underscore the profound influence of rich feature representations from SSL models, and improved geometric transformations with target TVs on the enhanced functionality of SI systems.
\end{abstract}


\vspace*{-3pt}
\section{Introduction}
Articulatory data refers to the positions and motion of different articulators in the vocal tract during speech. This data has shown to be critical in a number of speech applications, like speech therapy \cite{Fagel2008A3V}, and mental health assessment \cite{Siriwardena_SZ}. Articulatory data is obtained using different imaging techniques, like X-ray Microbeam (XRMB) \cite{Westbury1994a}, Electromagnetic Articulometry (EMA) \cite{Tiede2017} and real-time Magnetic Resonance Imaging (rt-MRI) \cite{Narayanan2004}. However, these methods are invasive, expensive, and can be dangerous under prolonged exposure \cite{Sivaraman_ASA}. Acoustic to articulatory Speech Inversion (SI) provides an alternative method of estimating the articulatory parameters from the acoustic signal. 

Deep Neural Networks (DNNs) have been shown to be effective SI systems \cite{siriwardena2022audio, siriwardena2023secret}. The performance of DNNs can be improved through better input and output feature space representation, and SI systems are no exception. In our previous works, we have shown that SI DNN models' performance can improve through better input representation through audio data augmentation \cite{siriwardena2022audio}, or incorporating source features \cite{siriwardena2023secret}.

Self-Supervised Learning (SSL) has been shown to be an effective method of improving DNN performance through the utilization of unlabeled data in learning speech representations\cite{hsu2021hubert, schneider2019wav2vec}. These representations have shown to be effective in Automatic Speech Recognition (ASR) systems \cite{conneau2020unsupervised}, speech separation and enhancement \cite{10094883}. Recent works have also shown that SSL speech representation has the capacity to improve the performance of SI models for EMA data \cite{berkely_ssl_si} outperforming the conventional acoustic features like Mel-frequency Cepstral Coefficients (MFCCs). Cho et al. \cite{berkely_ssl_si} have extensively evaluated the existing SSL speech representations for the SI task and have found that HuBERT based SSL speech representations \cite{hubert_ppr} works the best over both the other SSL features (eg. wav2vec2, tera \cite{tera_ppr}) and conventional acoustic features like MFCCs. 

Additionally, the analysis and prediction of raw articulatory data can be challenging. Raw articulatory data is represented in the absolute X-Y coordinates of different articulators, which is closely linked to the speaker's anatomy, leading to inter-speaker variability in pellet positions for the same sound. For that reason, quantifying vocal tract shape is best achieved by measuring the location and degree of these constrictions. These measurements are called Tract Variables (TVs) and can be achieved through geometric transformations of the raw articulatory parameters \cite{sivaraman2019unsupervised}. In a recent previous work, we have presented a novel geometric transformation which improved the performance of SI systems through better output feature space representation \cite{attia2023enhancing}.

In this work, we combine both approaches, by using HuBERT \cite{hsu2021hubert} SSL speech representation to improve the input representation. We also continue our previous work presented in \cite{attia2023enhancing} by proposing a new geometric transformation that enhances the performance of SI systems further. We show that using better input and output feature representations lead to better SI performance and more robust estimated TVs.

We begin by a description of the XRMB dataset in section \ref{sec:dataset}. We describe our novel TV transformation model in \ref{sec:transforms} and our experiments with SSL speech representations in \ref{sec:si_system}. Section \ref{sec:results} outlines the results of our experiments. We end-up with a conclusion and a discussion on our proposed future work in section \ref{sec:conclusion}.

\vspace*{-3pt}
\section{Articulatory Dataset}
\label{sec:dataset}

The original University of Wisconsin XRMB database \cite{Westbury1994a}, consists of naturally spoken isolated sentences and short paragraphs gathered from 32 male and 25 female participants. These speech recordings were accompanied by trajectory data obtained through X-ray microbeam cinematography of the midsagittal plane of the vocal tract. This cinematography tracked the movement of pellets placed on various articulators, including the upper lip (UL), lower lip (LL), tongue tip (T1), tongue blade (T2), tongue dorsum (T3), tongue root (T4), mandible incisor (MANi), and parasagittally placed mandible molar (MANm).

However, it's worth noting that some of the articulatory recordings in the database were flagged as mistracked. After removing these problematic samples, we were left with a total of 46 speakers (21 males and 25 females) and approximately 4 hours of speech data. In our recent work \cite{attia2023masked}, we reconstructed a large portion of the corrupted articulatory recordings. After adding the aforementioned reconstructed recordings to the original uncorrupted dataset, we were left with approximately 5.3 hours of speech data.

\begin{figure}
    \centering
    \includegraphics[width=\columnwidth]{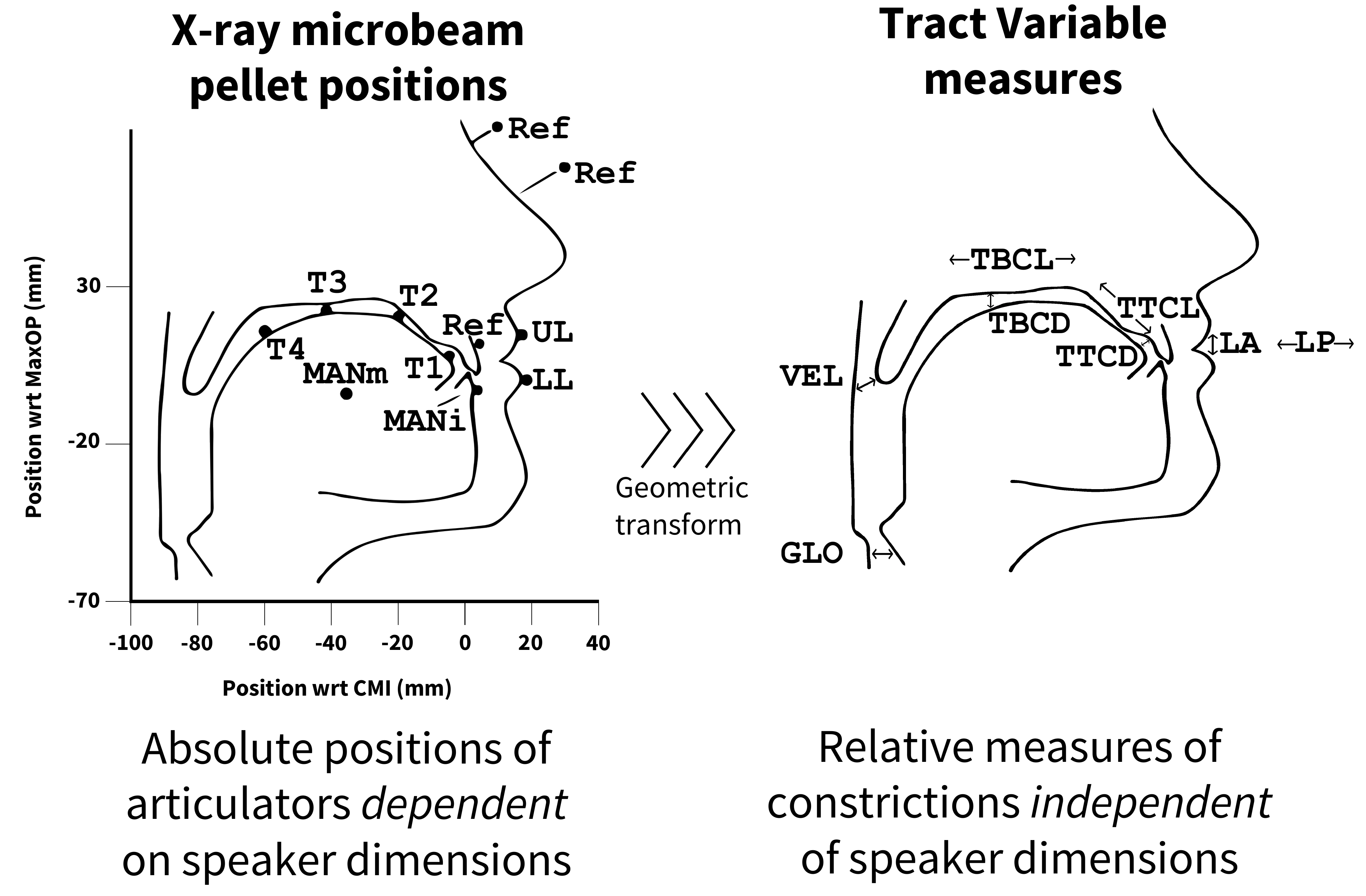}
    \caption{Pellet placement and TV definition in the XRMB dataset}

    \label{fig:xrmb}
\end{figure}

\vspace*{-3pt}
\section{Novel Tract Variable Transformations}
\label{sec:transforms}
As mentioned above, absolute X-Y coordinate representations of articulatory data is closely linked to speaker anatomy and leads to inter-speaker variability. To remedy this, the raw articulatory features are transformed into TVs using a geometric transformation. In this section, we outline a novel geometric transformation to extract TVs that are more closely related to the acoustic signal, which is a continuation of our work presented in \cite{attia2023enhancing}.

\subsection{Articulatory Model}
\subsubsection{Lips}
The lips are modeled using the UL and LL pellets. To describe the degree and location of lip constriction, we define two TVs, Lip Aperture (LA) and  Lip Protrusion (LP)  respectively. LA is defined as the euclidean distance between UL and LL. Unlike \cite{attia2023enhancing}, LP is defined as the horizontal offset of LL from the Y-axis instead of UL, which we empirically show that it leads to better SI performance. The origin of the X-Y plane is located at the tip of the maxillary incisors, and the X- axis is defined as the maxillary occlusal plane.

\begin{equation}
    LA[n] = || UL[n] - LL[n]||
\end{equation}
\begin{equation}
    LP[n] = LL_x[n]
\end{equation}

\subsubsection{Tongue Body}
\begin{figure}
    \centering
    \includegraphics[width = \columnwidth]{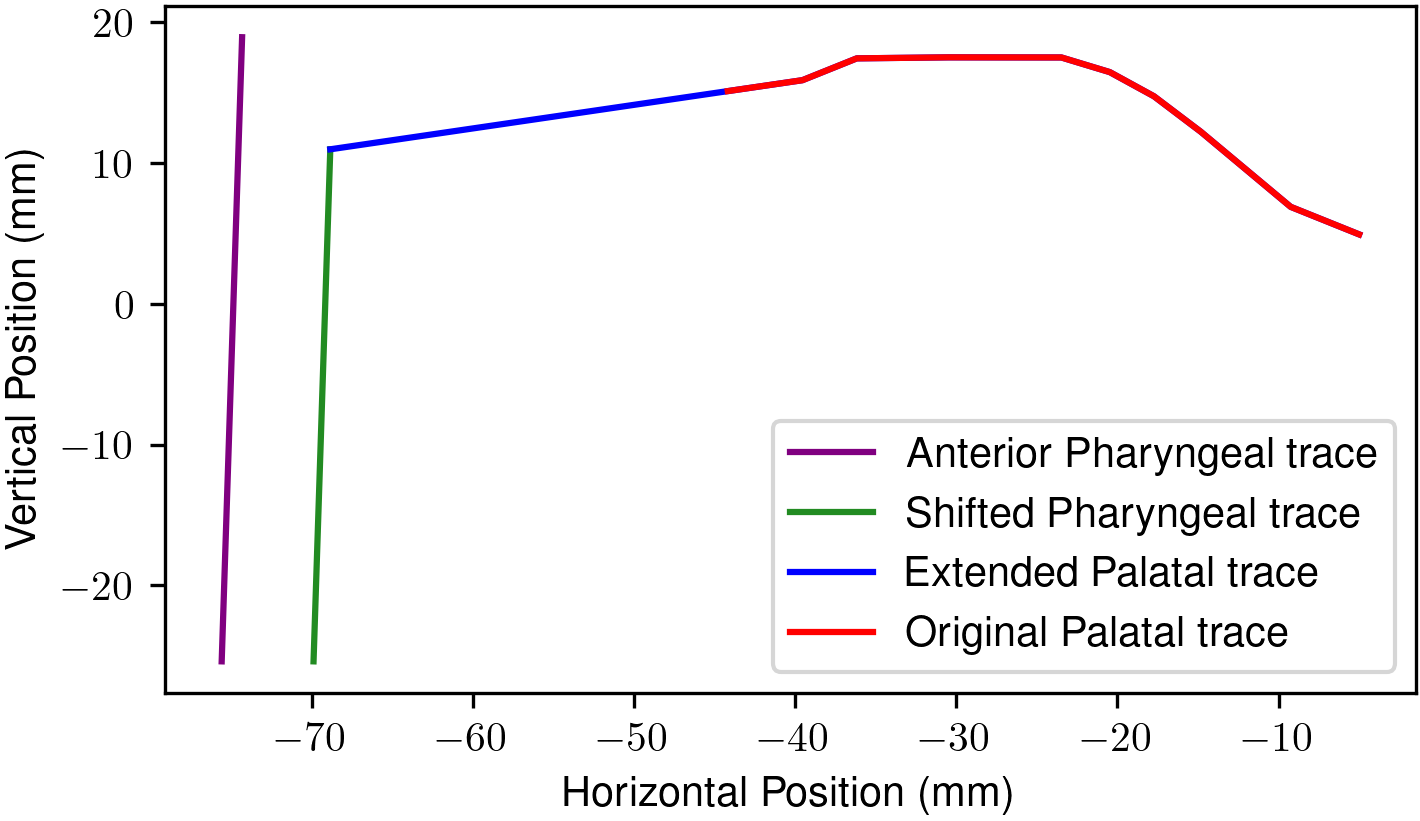}
    \caption{ Extended Palateal Trace With the Anterior Pharyn-
geal Wall For Speaker JW33}

    \label{fig:palatal_trace}
\end{figure}

The tongue body is modeled using a circle fitted through T2, T3 and T4. It's constriction can be described using two TVs, namely Tongue Body Constriction Location (TBCL) and Tongue Body Constriction Degree (TBCD). The constriction is measured relative to the extended palatal trace we introduced in \cite{attia2023enhancing}, which models the hard palate as well as the soft palate and the anterior pharyngeal wall. Figure \ref{fig:palatal_trace} shows the extended palatal trace.

TBCD is measured as the minimum Euclidean distance between the tongue body circle and the extended palatal trace. We update the definition of TBCL from the one introduced in \cite{attia2023enhancing} to be similar to the definition of LP. TBCL is defined as the horizontal offset of the point on the tongue body circle closest to the extended palatal trace, i.e. the point used in TBCD calculation from the Y-axis.
\begin{equation}
     TBCD = min_{p \in epal}[ min_{x \in TB_{circle}} || p - x||]
 \end{equation}
 \begin{equation}
  TBCL = -TB[argmin[TBCD]_x]
 \end{equation}
 
 where $epal$ is the extended palate trace, and \linebreak $TB[argmin[TBCD]]$ is the point on the tongue body closest to the palate trace
 
\subsubsection{Tongue Tip}

The tongue tip is modeled by the T1 pallet. It's constriction can be described by two TVs, Tongue Tip Constriction Location (TTCL) and Tongue Tip Constriction Degree (TTCD). Similar to TBCD and TBCL, TTCD is defined as the minimum Euclidean distance between T1 and the extended pallatal trace, and TTCL is the horizontal offset of T1 from the Y-axis.

\begin{equation}
     TTCD = min_{p \in epal}[ || p - T1||]
 \end{equation}
 \begin{equation}
  TBCL = -T1_x
 \end{equation}

\begin{table*}[ht]

    \caption{PPMC between predicted and ground truth TVs for SI systems trained on datasets according to each geometric transformation model, with the MFCCs and HuBERT input features.\\}
  \centering
  
  \resizebox{2\columnwidth}{!}{
  \large
  \begin{tabular}{|c |c|c | c| c| c| c|c| c|}
    \hline
   \rule{0pt}{3ex}
   
  \begin{tabular}{@{}c@{}}\textbf{\large{Transformation}} \\ \textbf{\large{Model}}\end{tabular}&\textbf{Training Dataset}&\textbf{LA} &  \textbf{LP} & \textbf{TBCL} & \textbf{TBCD} & \textbf{TTCL} & \textbf{TTCD} &  \textbf{Average}\\
  \hline
 
      \multicolumn{9}{|c|}{\textbf{MFCC Input Features}}\\
      \hline
       \rule{0pt}{3ex}
  \textbf{Baseline} & Small Dataset& 0.8679 &0.5902 &0.7424 &0.7801&0.5971&0.8934&0.7452\\
  \textbf{Proposed} &Small Dataset &0.8603 &0.7104 &0.7426&0.7754&\textbf{0.7422}&0.8981&0.7881\\
  
  \textbf{Proposed} & Extended Dataset &\textbf{0.8697} &	\textbf{0.7250} &	\textbf{0.7508} &	\textbf{0.7847} &	0.7407 &	\textbf{0.9019} &	\textbf{0.7955}\\
  \hline
  \multicolumn{9}{|c|}{\textbf{HuBERT Input Features}}\\
  \hline
  
  \textbf{Proposed} & Small Dataset &0.8779&\textbf{0.7243}&\textbf{0.7430}&0.8089&0.7865&\textbf{0.9248}&0.8109\\
  \textbf{Proposed} & Extended Dataset &\textbf{0.8902}&0.7142&0.7361&\textbf{0.8180}&\textbf{0.8032}&0.9229&\textbf{0.8141}\\
  \hline
  \end{tabular}}
   \label{table: transformation}\\
   
\end{table*}

\section{Speech Inversion Model Architectures}
\label{sec:si_system}

This section describes the experimented SI model architectures and details on model training.

\subsection{SI Architecture with HuBERT features}

\label{ssec:si_hubert}
SSL speech representations when used in the SI task with EMA data have shown to outperform the conventional acoustic features (eg. Mel-spectrograms, Mel-frequency Cepstral Coefficients (MFCCs)) \cite{berkely_ssl_si}. Here the SSL representations only need to be fine-tuned for the down stream task of speech inversion and can be expected to generalize better even with limited ground-truth articulatory data. Based on the previous work in \cite{berkely_ssl_si} for using SSL features for the SI task with EMA data, we explored the idea of using HuBERT SSL features \cite{hubert_ppr} as the input acoustic representation to train our best performing Bidirectional Gated Recurrent Unit (BiGRNN) SI architecture.  

We used the HuBERT-large model pre-trained with the Librilight dataset (60,000h) to extract the HuBERT speech embeddings. All the audio files (sampled with 16 KHz) are first segmented to 2 second long segments and the shorter ones are zero padded at the end. The HuBERT embeddings are then extracted from the 2 second long segments using the speechbrain open-source AI toolkit \cite{speechbrain}. The HuBERT embeddings are sampled at 50 Hz and have a dimensionality of 1024. 

We used the BiGRNN SI system proposed in \cite{siriwardena2022audio}, and adapted the input layer to match the input dimensionality of the HuBERT embeddings.

\subsection{SI Architecture with MFCC features}

\label{ssec:si_mfcc}
We trained the same SI system architecture used in \cite{attia2023enhancing} which is identical to that discussed in section \ref{ssec:si_hubert} with the only difference being 13 MFCCs used as the input acoustic feature. The MFCCs were
extracted using a 20ms Hamming analysis window with a 10ms frame shift. The MFCCs are also utterance wise normalized (z-normalized) prior to model training.

\subsection{Model Training}

Both the SI architectures described above were trained in similar fashion. The input XRMB dataset was first divided into training, development, and testing sets, so that the training set has utterances from 36 speakers and the development and testing sets have 5 speakers each (3 males, 2 females). None of the training, development and testing sets have overlapping speakers and hence all the models were trained in a ‘speaker-independent’ fashion. All the models were implemented with a TensorFlow-Keras machine learning framework. ADAM optimizer with a starting learning rate of 1e-3 and an exponential learning rate scheduler was used. Both the models with HuBERT and MFCCs were trained with an early stopping criteria (patience=5) monitoring the ‘validation loss’ on the development set. To choose the best starting ‘learning rate’, we did a grid search on [1e-3, 3e-4, 1e-4], whereas to choose the training batch size, we did a similar grid search on [16,32,64,128]. Based on the validation loss, 1e-3 and 32 were chosen as the learning rate and batch size, respectively, for the model with HuBERT features and 1e-3 and 64 for the model with MFCCs.

\vspace*{-3pt}
\section{Results}

\label{sec:results}

\subsection{TV Transformations}

\label{sec:tv_transforms}
 In this subsection, we evaluate our new transformation model and compare it to the baseline, which is our previous model introduced in \cite{attia2023enhancing}. We evaluate the Transformation models by training the same DNN SI model on the same dataset not including any reconstructed files, transformed, which we call the `small dataset', according to each respective geometric transformation. We also evaluate the SI model when trained on the entire available training data including reconstructed files, which we call the `extended dataset'. We argue that the better the SI performance, the more closely related the resulting TVs are to the acoustic input signal. We evaluate the SI model based on the Pearson Product Moment Correlation (PPMC) between the predicted and ground truth TVs. 

The first part of Table \ref{table: transformation} shows the performance of the SI with MFCC input features. Our proposed model outperforms the baseline on average, by 4.3\% on the small dataset, with noticeable improvement in LP and TTCL. Training on the extended dataset also improves performance across the board over the small dataset. Overall, the combination of better transformation and more data has improved the performance of the SI system by 5.03\%. 

However, it is worth noting that improving the TV geometric transformation model was more effective than increasing the size of the training data. This highlights the importance of having better output feature space representation.

\subsection{SSL features with new TV transformations}

In this subsection, we discuss the effect of using HuBERT speech representation as in input to the SI system juxtaposed with MFCCs. The two model architectures are discussed in section \ref{ssec:si_hubert} and section \ref{ssec:si_mfcc}.

Training on the small dataset, HuBERT  representation lead to a tangible improvement in the tongue TVs, namely TBCL, TBCD, TTCL and TTCD, with slight improvement in LA and LP. On average, using HuBERT representations lead to a 2.3\% improvement in PPMC scores. 

Training on a extended dataset leads to some improvement, although not significant when compared to improving input representations. On average adding more training data increases PPMC by 0.32\%. This again highlights the effect of input representation, which was more effective than increasing the training data size. Overall, by combining better input and output representations, along with including more data, we were able to improve the PPMC score by from 0.7452 to 0.8141, a 6.9\% improvement.

\subsection{Estimated TVs with best performing SI systems}

\begin{figure}[th]
    \centering
    \includegraphics[width=\linewidth, height = 88mm]{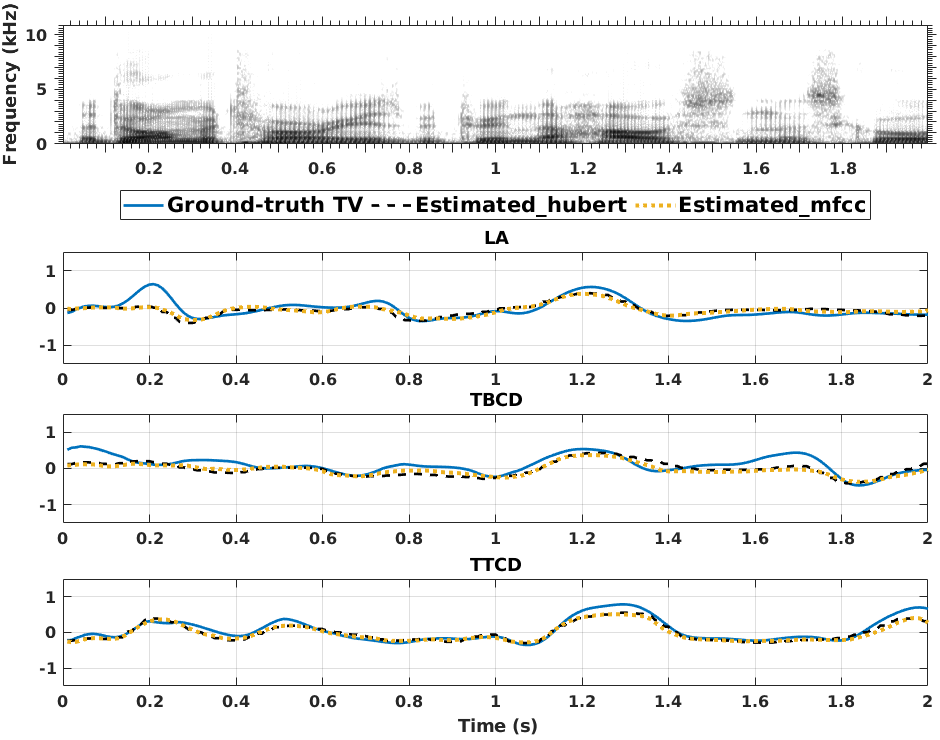}
    \caption{LA and constriction degree TVs for the utterance ‘The dormitory is between the house and the school’ estimated by the model trained with HuBERT embeddings (estimated\_hubert) and the model trained with MFCCs (estimated\_mfcc). Solid blue Line - ground truth, black dotted line - predictions by the HuBERT based model, yellow dotted Line - predictions by MFCC based model.}

    \label{fig:tv_plots}
\end{figure}

Figure \ref{fig:tv_plots} shows the estimated LA and constriction degree TVs for an utterance in the test set, by the two SI systems trained with HuBERT and MFCC features. Both the systems have been trained with the `extended dataset'. As seen in the figure, the differences between the TV estimates by the two models are subtle. But consistent with the PPMC scores in Table \ref{table: transformation}, it can be seen that both LA and TTCD TVs are estimated better with the HuBERT based model compared to the model trained with MFCCs. It can also be seen that the TBCD is better estimated by the model trained with MFCCs compared to the HuBERT based model.

\vspace*{-3pt}
\section{Conclusion and Future Work}
\vspace*{-2pt}

\label{sec:conclusion}

In this paper, we propose a new geometric transformation to obtain TVs from raw XRMB pellets. We show that our novel TV transformation improves the performance of SI model over the baseline, which can be attributed to better relation between the resulting TVs and the acoustic signal. We also further improve the performance of the SI system by using HuBERT speech representation as the input to the SI model. Our findings highlight the importance of efficient input and output feature space representations.

In \cite{attia2023enhancing}, we highlighted some of the limitations of the TV transformation model we proposed in that paper. In this paper, we tackled a majority of these limitations. However, we still lag behind the transformation proposed in \cite{sivaraman2019unsupervised} with respect to TBCL, even though we achieve better PPMC on average. This can be attributed to the fact that even though we extended the palatal trace towards the anterior pharyngeal wall, we didn't extend the tongue body circle beyond the T4. Even though this model improved the representation of TBCD, which means it gave a good estimate for the degree of the constriction, not extending the tongue body circle might lead to inaccurate estimation of the location of the constriction, which in turn would lead to lower correlation between TTCL and the acoustic signal. We intend to tackle this problem in future work.

\vspace{12pt}


\vspace*{-3pt}
\begin{thebibliography}{00}
\bibitem{Mitra, Vikramjit and Espy-Wilson, Carol Y. and Saltzman, Elliot and Goldstein, Louis} Mitra, Vikramjit and Espy-Wilson, Carol Y. and Saltzman, Elliot and Goldstein, Louis, "Articulatory Information for Noise Robust Speech Recognition," IEEE Transactions on Audio, Speech, and Language Processing, vol. 19, no. 7, pp. 1913-1924, 2011.

\bibitem{Ling, Zhen-Hua and Richmond, Korin and Yamagishi, Junichi} Ling, Zhen-Hua and Richmond, Korin and Yamagishi, Junichi, "Articulatory Control of HMM-Based Parametric Speech Synthesis Using Feature-Space-Switched Multiple Regression," IEEE Transactions on Audio, Speech, and Language Processing, vol. 21, no. 1, pp. 207-219, 2013.


\bibitem{Siriwardena, Yashish M and Espy-Wilson, Carol and Kitchen, Chris and Kelly, Deanna L} Siriwardena, Yashish M and Espy-Wilson, Carol and Kitchen, Chris and Kelly, Deanna L, "Multimodal Approach for Assessing Neuromotor Coordination in Schizophrenia Using Convolutional Neural Networks," Proceedings of the 2021 International Conference on Multimodal Interaction,pp. 768--772, 2021.
\bibitem{Carol Espy-Wilson and Adam C. Lammert and Nadee Seneviratne and Thomas F. Quatieri} Carol Espy-Wilson and Adam C. Lammert and Nadee Seneviratne and Thomas F. Quatieri, "Assessing Neuromotor Coordination in Depression Using Inverted Vocal Tract Variables," Proc. Interspeech 2019,1448--1452, 2019. [Online]. Available: 10.21437/Interspeech.2019-1815
\bibitem{Westbury, John R} Westbury, John R, "Speech Production Database User ' S Handbook," IEEE Personal Communications - IEEE Pers. Commun., vol. 0, no. June

\bibitem{Mitra2011_new} Mitra, Vikramjit and Espy-Wilson, Carol Y. and Saltzman, Elliot and Goldstein, Louis, "Articulatory Information for Noise Robust Speech Recognition," IEEE Transactions on Audio, Speech, and Language Processing, vol. 19, no. 7, pp. 1913-1924, 2011.
\bibitem{speech_synthesis_1} Ling, Zhen-Hua and Richmond, Korin and Yamagishi, Junichi, "Articulatory Control of HMM-Based Parametric Speech Synthesis Using Feature-Space-Switched Multiple Regression," IEEE Transactions on Audio, Speech, and Language Processing, vol. 21, no. 1, pp. 207-219, 2013. 
\bibitem{Fagel2008A3V} Ling, Zhen-Hua and Richmond, Korin and Yamagishi, Junichi, "Articulatory Control of HMM-Based Parametric Speech Synthesis Using Feature-Space-Switched Multiple Regression," IEEE Transactions on Audio, Speech, and Language Processing, vol. 21, no. 1, pp. 207-219, 2013.
\bibitem{Siriwardena_SZ} Siriwardena, Yashish M. and Espy-Wilson, Carol and Kitchen, Chris and Kelly, Deanna L., "Multimodal Approach for Assessing Neuromotor Coordination in Schizophrenia Using Convolutional Neural Networks," Proceedings of the 2021 International Conference on Multimodal Interaction,pp. 768–772, 2021. 
\bibitem{espywilson19_interspeech} Carol Espy-Wilson and Adam C. Lammert and Nadee Seneviratne and Thomas F. Quatieri, "Assessing Neuromotor Coordination in Depression Using Inverted Vocal Tract Variables," Proc. Interspeech 2019,pp. 1448--1452, 2019. [Online]. Available: 10.21437/Interspeech.2019-1815
\bibitem{Westbury1994a} Westbury, John R, "Speech Production Database User ' S Handbook," IEEE Personal Communications - IEEE Pers. Commun., vol. 0, no. June, 
\bibitem{Tiede2017} Tiede,Mark  and Espy-Wilson,Carol Y.  and Goldenberg,Dolly  and Mitra,Vikramjit  and Nam,Hosung  and Sivaraman,Ganesh, "Quantifying kinematic aspects of reduction in a contrasting rate production task," The Journal of the Acoustical Society of America, vol. 141, no. 5, pp. 3580-3580, 2017. [Online]. Available: 10.1121/1.4987629
\bibitem{Narayanan2004} Narayanan, Shrikanth and Nayak, Krishna and Lee, Sungbok and Sethy, Abhinav and Byrd, Dani, "An approach to real-time magnetic resonance imaging for speech production," The Journal of the Acoustical Society of America, vol. 115, no. 4, pp. 1771--1776, 2004.
\bibitem{Sivaraman_ASA} Sivaraman,Ganesh  and Mitra,Vikramjit  and Nam,Hosung  and Tiede,Mark  and Espy-Wilson,Carol, "Unsupervised speaker adaptation for speaker independent acoustic to articulatory speech inversion," The Journal of the Acoustical Society of America, vol. 146, no. 1, pp. 316-329, 2019. 
\bibitem{siriwardena2022audio} Y. M. Siriwardena, A. A. Attia, G. Sivaraman and C. Espy-Wilson, "Audio Data Augmentation for Acoustic-to-Articulatory Speech Inversion," 2023 31st European Signal Processing Conference (EUSIPCO), Helsinki, Finland, 2023, pp. 301-305, doi: 10.23919/EUSIPCO58844.2023.10290069,
\bibitem{siriwardena2023secret} Siriwardena, Yashish M and Espy-Wilson, Carol, "The Secret Source: Incorporating Source Features to Improve Acoustic-To-Articulatory Speech Inversion," ICASSP 2023-2023 IEEE International Conference on Acoustics, Speech and Signal Processing (ICASSP),pp. 1--5, 2023. 
\bibitem{siriwardena2022learning} Siriwardena, Y.M., Espy-Wilson, C., Shamma, S. (2023) Learning to Compute the Articulatory Representations of Speech with the MIRRORNET. Proc. INTERSPEECH 2023, 5137-5141, doi: 10.21437/Interspeech.2023-562
\bibitem{sivaraman2016vocal} Sivaraman, Ganesh and Mitra, Vikramjit and Nam, Hosung and Tiede, Mark K and Espy-Wilson, Carol Y, "Vocal Tract Length Normalization for Speaker Independent Acoustic-to-Articulatory Speech Inversion.," INTERSPEECH,pp. 455--459, 2016. 
\bibitem{seneviratne2019multi} Seneviratne, Nadee and Sivaraman, Ganesh and Espy-Wilson, Carol Y, "Multi-Corpus Acoustic-to-Articulatory Speech Inversion.," Interspeech,pp. 859--863, 2019. 
\bibitem{attia2023enhancing} Attia, Ahmed Adel and Tiede, Mark and Espy-Wilson, Carol Y, "Enhancing Speech Articulation Analysis using a Geometric Transformation of the X-ray Microbeam Dataset," arXiv preprint arXiv:2305.10775,
\bibitem{hsu2021hubert} Hsu, Wei-Ning and Bolte, Benjamin and Tsai, Yao-Hung Hubert and Lakhotia, Kushal and Salakhutdinov, Ruslan and Mohamed, Abdelrahman, "Hubert: Self-supervised speech representation learning by masked prediction of hidden units," IEEE/ACM Transactions on Audio, Speech, and Language Processing, vol. 29, pp. 3451--3460, 2021. 
\bibitem{schneider2019wav2vec} Schneider, Steffen and Baevski, Alexei and Collobert, Ronan and Auli, Michael, "wav2vec: Unsupervised pre-training for speech recognition," arXiv preprint arXiv:1904.05862,
\bibitem{baevski2020wav2vec} Baevski, Alexei and Zhou, Yuhao and Mohamed, Abdelrahman and Auli, Michael, "wav2vec 2.0: A framework for self-supervised learning of speech representations," Advances in neural information processing systems, vol. 33, pp. 12449--12460, 2020. 
\bibitem{conneau2020unsupervised} Conneau, Alexis and Baevski, Alexei and Collobert, Ronan and Mohamed, Abdelrahman and Auli, Michael, "Unsupervised cross-lingual representation learning for speech recognition," arXiv preprint arXiv:2006.13979,
\bibitem{fan2020exploring} Fan, Zhiyun and Li, Meng and Zhou, Shiyu and Xu, Bo, "Exploring wav2vec 2.0 on speaker verification and language identification," arXiv preprint arXiv:2012.06185,
\bibitem{10094883} Wang, Tianrui and Chen, Xie and Chen, Zhuo and Yu, Shu and Zhu, Weibin, "An Adapter Based Multi-Label Pre-Training for Speech Separation and Enhancement," ICASSP 2023 - 2023 IEEE International Conference on Acoustics, Speech and Signal Processing (ICASSP), vol. , no. , pp. 1-5, 2023. [Online]. Available: 10.1109/ICASSP49357.2023.10094883
\bibitem{10094703} Udupa, Sathvik and C, Siddarth and Ghosh, Prasanta Kumar, "Improved Acoustic-to-Articulatory Inversion Using Representations from Pretrained Self-Supervised Learning Models," ICASSP 2023 - 2023 IEEE International Conference on Acoustics, Speech and Signal Processing (ICASSP), vol. , no. , pp. 1-5, 2023. [Online]. Available: 10.1109/ICASSP49357.2023.10094703
\bibitem{MCGOWAN199419} McGowan, Richard S, "Recovering articulatory movement from formant frequency trajectories using task dynamics and a genetic algorithm: Preliminary model tests," Speech Communication, vol. 14, no. 1, pp. 19--48, 1994. 
\bibitem{attia2023masked} Attia, Ahmed Adel and Espy-Wilson, Carol Y, "Masked Autoencoders are Articulatory Learners," ICASSP 2023-2023 IEEE International Conference on Acoustics, Speech and Signal Processing (ICASSP),pp. 1--5, 2023. 
\bibitem{sivaraman2019unsupervised} Sivaraman, Ganesh and Mitra, Vikramjit and Nam, Hosung and Tiede, Mark and Espy-Wilson, Carol, "Unsupervised speaker adaptation for speaker independent acoustic to articulatory speech inversion," The Journal of the Acoustical Society of America, vol. 146, no. 1, pp. 316--329, 2019. 
\bibitem{schneider19_interspeech} Steffen Schneider and Alexei Baevski and Ronan Collobert and Michael Auli, "wav2vec: Unsupervised Pre-Training for Speech Recognition," Proc. Interspeech 2019,pp. 3465--3469, 2019. [Online]. Available: 10.21437/Interspeech.2019-1873
\bibitem{mockingjay} Liu, Andy T. and Yang, Shu-wen and Chi, Po-Han and Hsu, Po-chun and Lee, Hung-yi, "Mockingjay: Unsupervised Speech Representation Learning with Deep Bidirectional Transformer Encoders," ICASSP 2020 - 2020 IEEE International Conference on Acoustics, Speech and Signal Processing (ICASSP), vol. , no. , pp. 6419-6423, 2020. 
\bibitem{hubert_ppr} Hsu, Wei-Ning and Bolte, Benjamin and Tsai, Yao-Hung Hubert and Lakhotia, Kushal and Salakhutdinov, Ruslan and Mohamed, Abdelrahman, "HuBERT: Self-Supervised Speech Representation Learning by Masked Prediction of Hidden Units," IEEE/ACM Trans. Audio, Speech and Lang. Proc., vol. 29, pp. 3451–3460, 2021. [Online]. Available: 10.1109/TASLP.2021.3122291
\bibitem{berkely_ssl_si} Cho, Cheol Jun and Wu, Peter and Mohamed, Abdelrahman and Anumanchipalli, Gopala K., "Evidence of Vocal Tract Articulation in Self-Supervised Learning of Speech," ICASSP 2023 - 2023 IEEE International Conference on Acoustics, Speech and Signal Processing (ICASSP), vol. , no. , pp. 1-5, 2023. 
\bibitem{tera_ppr} Liu, Andy T. and Li, Shang-Wen and Lee, Hung-yi, "TERA: Self-Supervised Learning of Transformer Encoder Representation for Speech," IEEE/ACM Trans. Audio, Speech and Lang. Proc., vol. 29, pp. 2351–2366, 2021. [Online]. Available: 10.1109/TASLP.2021.3095662
\bibitem{speechbrain} Ravanelli, Mirco and Parcollet, Titouan and Plantinga, Peter and Rouhe, Aku and Cornell, Samuele and Lugosch, Loren and Subakan, Cem and Dawalatabad, Nauman and Heba, Abdelwahab and Zhong, Jianyuan and others, "SpeechBrain: A general-purpose speech toolkit," arXiv preprint arXiv:2106.04624,
\end{thebibliography}
\end{document}